\documentclass[12pt]{iopart}
\usepackage[square,comma,numbers,sort&compress]{natbib}

\def\dd{{\rm d}}
\def\ie{{\em{i.e.}}}

\def\eg{{\em{e.g.}}}

\newcommand{\ve}{\varepsilon}
\newcommand{\vare}{\varepsilon_{\sf e}}
\newcommand{\varm}{\varepsilon_{\sf m}}
\newcommand{\vari}{\varepsilon_{\sf i}}

\begin{document}

\paper{Analogy between dielectric relaxation and dielectric mixtures: Application of the spectral density representation}
\author{Enis Tuncer}
\ead{enis.tuncer@physics.org}
\date{\today}

\begin{abstract}
The spectral representation is an effecient tool to explore electrical properties of material mixtures. It separates the contributions of geometrical topology and intrinsic properties of the constituents in the system. The aim of this paper is to derive an expression for the spectral density representation, which favors dielectric relaxation phenomenon. This unfamiliar form is distinct in a way that the existing dielectric relaxation models and data analysis tools can be employed for extracting the spectral density function of a given system. 
\end{abstract}


Electrical properties of material mixtures have attracted researchers in academia and industry to seek a relation between overall composite properties and intrinsic properties of the constituents and their spatial arrangement inside the mixture~\cite{Landauer1978}. 
Bergman~\cite{bergman1982} 
has proposed a mathematical way for representing the effective dielectric permittivity $\varepsilon_{\sf e}$ of a binary mixture as a function of permittivities of its constituents, $\varepsilon_{\sf m}$ and $\varepsilon_{\sf i}$, and an integral equation, which includes the geometrical contributions. This theory is called {\em the spectral density representation}. Milton~\cite{Milton1981} 
gave the corrections, whereas,  Golden and Papanicolaou~\cite{Golden1} presented the rigorous derivation for the spectral representation. 
The permittivity $\vare$ of a mixture  is expressed as follows in the spectral representation~\cite{GhoshFuchs} 
\begin{eqnarray}
  \label{eq:generalwiener_e}
  \vare&=&\varm\{ 1+qA\left({\vari}{\varm}^{-1}-1\right)+\nonumber\\ 
    && {\int}_{0^{+}}^{1}{q{\sf G}(x)}[{\left({\displaystyle \vari}{\displaystyle \varm}^{-1}-1\right)^{-1}+x}]^{-1}\dd x \}  
\end{eqnarray}
where $q$ and $x$ are the concentration of inclusions and spectral parameter, respectively, and  ${\sf G}(x)$ is the spectral density function. The constant $A$ is related to the percolation strength, which includes the contribution of ${\sf G}(0)$. Eq.~(\ref{eq:generalwiener_e}) can be expanded by substituting $\varepsilon_i-\varepsilon_j\equiv\Delta_{ij}$, then Eq.~(\ref{eq:generalwiener_e}) becomes,
\begin{eqnarray}
  \label{eq:A2}
  {\Delta_{\sf em}}{\varm}^{-1}&=&q\ A {\Delta_{\sf im}}{\varm}^{-1}+\nonumber\\
  &&\int_{0^{+}}^1{q{\sf G}(x)\Delta_{\sf im}}({\varm+\Delta_{\sf im}x})^{-1}\dd x
\end{eqnarray}
Now, multiplying both sides with $\varm$ and letting $\Delta_{\sf em}/\Delta_{\sf im}\equiv\xi$ and $qA\equiv=\xi_s$, we obtain 
\begin{eqnarray}
  \label{eq:spectral}
  \xi= \xi_s+q\int_{0^+}^1 {{\sf G}(x)}({1+\varepsilon_{\sf m}^{-1}\Delta_{\sf im}x})^{-1} \dd x
\end{eqnarray}
Here, we call $\xi$ the `scaled permittivity' and $\xi_s$ the percolation strength as defined earlier. The mathematical properties of ${\sf G}$, $q$ and $\xi_s$ are expressed in the literature\cite{GhoshFuchs}; Here $\xi_s+q\int_{0^{+}}^1 {\sf G}(x) \dd x =1$ and $\int_{0^+}^{1} x {\sf G}(x) \dd x =(1-q){d}^{-1}$, where $d$ is the dimensionality of the system. In view of Eq.~(\ref{eq:spectral}), the dielectric relaxation expression for a process with single relaxation time $\tau$ is expressed as~\cite{Debye1945,Tuncer2000b}
\begin{eqnarray}
  \label{eq:debye}
  \ve(\imath\omega)= \ve_\infty+\Delta\ve({1+\imath\omega\tau})^{-1}
\end{eqnarray}
where $\ve_\infty$, $\Delta\ve$ and $\omega$ are the permittivity at optical frequencies $(\omega\rightarrow\infty)$, dielectric strength and angular frequency respectively.   If there exist a continuous relaxation with a distribution, Eq.~(\ref{eq:debye}) becomes
 \begin{eqnarray}
  \label{eq:drt}
  \ve(\imath\omega)= \ve_\infty+\Delta\ve\int_0^\infty {{\sf G}(\tau) }({1+\imath\omega\tau})^{-1} \dd \tau.
\end{eqnarray}
where ${\sf G}(\tau)$ is the distribution function of the relaxation times, and the static dielectric permittivity $\ve_s$ is defined as $\ve_\infty + \Delta\ve\int_0^\infty {\sf G}(\tau)\dd \tau=\ve_s$. Observe the similarities between Eqs.~(\ref{eq:spectral}) and (\ref{eq:drt}). Although the resemblence between two equations looks superficial, when Claussius-Mossotti \cite{Debye1945,Onsager} expression is taken into consideration, then the frequency dependent properties of dielectrics can infact be written as a dynamic response of dipole units embedded in a background medium $\varepsilon_{\sf m}$, concluding a dielectric mixture approach with inclusions as dipole units. Consequenctly, interfacial polarization observed in dielectric mixtures can be used to understand the structure-property relationship in physics of dielectrics.

There are couple of expressions that are extensively used in the dielectric data analysis, \eg\ Havriliak-Negami~\cite{HN}, Davidson-Cole~\cite{CD}, Cole-Cole~\cite{CC}, which have known distribution functions ${\sf G}(\tau)$~\cite{Tuncer2000b}. In addition, there exists a vast literature on dielectric data analysis~\cite{Jonscher1983,Tuncer2000b}. 
The total polarizability of a material is given as $\varepsilon(\omega\rightarrow0)$, therefore, $\ve_\infty +\Delta\ve=\ve_s$, which is actually similar to the definitions and properties of ${\sf G}(x)$ and $\xi_s$, which can be obtained by converting polarizations to fractional polarizations, which results by dividing both sides by $\ve_s$. As a consequence, we can complete the derivation and the hypothesis about the relation between the dielectric relaxation and two-component dielectric mixtures as the relationship between various parameters in both representations, $\varepsilon_\infty\ve_s^{-1}\Leftrightarrow\xi_s$, $\Delta\varepsilon\ve_s^{-1}\Leftrightarrow q$ and ${\sf G}(x)  \Leftrightarrow {\sf G}(\tau)$ and $\omega\Leftrightarrow\varpi\equiv\varm^{-1}\Delta_{\sf im}$, where $\varpi$ is the scaled frequency of the mixture system. 

Due to the similarities between Eqs.~(\ref{eq:spectral}) and (\ref{eq:drt}), and the commutability of various parameters, one can express a general `scaled' dielectric mixture formula as in the dielectric dispersions, \ie\ the Havriliak-Negami expression~\cite{HN},
 \begin{eqnarray}
  \label{eq:1}
  \xi(\varpi)=\xi_s+{q}[1+(\varpi x)^{\alpha}]^{-\beta}
\end{eqnarray}
As a result the dielectric permittivity $\vare$ of the mixture then  becomes
\begin{eqnarray}
  \label{eq:newe}
  \vare=\varm+\Delta_{\sf im}\{\xi_s+{q}[1+(\varpi x)^{\alpha}]^{-\beta}\}.
\end{eqnarray}
Yet the scaled permittivity notation can somewhat be used to calculate the time-domain dielectric mixture relations by using an inverse transform as in the case of dielectric relation theory, \ie\ the relationship between the response function and the dielectric susceptibility. As note, one more resemblance between the two representations expressed by the Havriliak-Negami expression is that it yields the Maxwell Garnett equation~\cite{Maxwell_Garnett}, which formulates the dielectric permittivity of a mixture with spherical inclusions ($d=3$), while in the dielectric relaxation form it  results in the simple Debye relaxation in Eq.~(\ref{eq:debye}), when $\alpha=\beta=1$, $\xi_s\approx0$ and $x=(1-q)d^{-1}$. Similarly for Claussius-Mossotti expression, $\alpha=\beta=1$, and  the spectral parameter localizes at $3^{-1}$; and $q$ becomes the number density of dipoles.

In this short paper, it is despised that the dielectric relaxation phenomenon and expressions developed thereof for analyzing dielectric dispersions could be employed to investigate the topological description or spectral density functions of two-component composites. Once the data are transformed to the scaled permittivity notation as shown in Eqs.~(\ref{eq:spectral}) or (\ref{eq:1}), the existing dielectric relaxation data analysis tools infact can readily be applied, there exist many dispersion expressions in the literature. Finally, it can be inferred that even pure one-component materials can also be expressed in terms of scaled permittivity notation, wherein, the smallest parts of the material are embedded in vacuum $\varm=1$. In such case the spectral density function would be indicative of the structure of the material and its parts, which in turn can be valuable to calculate the local fields~\cite{Onsager} and interaction energies~\cite{BarreraPRB1995}.
~\\

I thank Mr Rajeev Singh for fruitful discussion and  for valuable comments while preparing this paper. 


\begin{thebibliography}{24}
\expandafter\ifx\csname natexlab\endcsname\relax\def\natexlab#1{#1}\fi
\expandafter\ifx\csname bibnamefont\endcsname\relax
  \def\bibnamefont#1{#1}\fi
\expandafter\ifx\csname bibfnamefont\endcsname\relax
  \def\bibfnamefont#1{#1}\fi
\expandafter\ifx\csname citenamefont\endcsname\relax
  \def\citenamefont#1{#1}\fi
\expandafter\ifx\csname url\endcsname\relax
  \def\url#1{\texttt{#1}}\fi
\expandafter\ifx\csname urlprefix\endcsname\relax\def\urlprefix{URL }\fi
\providecommand{\bibinfo}[2]{#2}
\providecommand{\eprint}[2][]{\url{#2}}

\bibitem[{\citenamefont{Landauer}(1978)}]{Landauer1978}
\bibinfo{author}{\bibfnamefont{R.}~\bibnamefont{Landauer}}, in
  \emph{\bibinfo{booktitle}{Electrical Transport and Optical properties of
  Inhomogeneous Media}}, edited by \bibinfo{editor}{\bibfnamefont{J.~C.}
  \bibnamefont{Garland}} \bibnamefont{and}
  \bibinfo{editor}{\bibfnamefont{D.~B.} \bibnamefont{Tanner}}
  (\bibinfo{publisher}{American Institute of Physics}, \bibinfo{address}{New
  York}, \bibinfo{year}{1978}), vol.~\bibinfo{volume}{40} of
  \emph{\bibinfo{series}{AIP Conference Proceedings}}, pp.
  \bibinfo{pages}{2--43};
%
\bibinfo{editor}{\bibfnamefont{L.~V.} \bibnamefont{Keldysh}},
  \bibinfo{editor}{\bibfnamefont{D.~A.} \bibnamefont{Kirzhnitz}},
  \bibnamefont{and} \bibinfo{editor}{\bibfnamefont{A.~A.}
  \bibnamefont{Maradudin}}, eds., \emph{\bibinfo{title}{The Dielectric Function
  of Condensed Systems}} (\bibinfo{publisher}{Elsevier Science Publisher B.V.},
  \bibinfo{address}{Amsterdam}, \bibinfo{year}{1989});
\bibinfo{editor}{\bibfnamefont{A.}~\bibnamefont{Priou}}, ed.,
  \emph{\bibinfo{title}{Progress in Electromagnetics Research}}, Dielectric
  Properties of Heterogeneous Materials (\bibinfo{publisher}{Elsevier},
  \bibinfo{address}{New York}, \bibinfo{year}{1992});
\bibinfo{author}{\bibfnamefont{D.~J.} \bibnamefont{Bergman}} \bibnamefont{and}
  \bibinfo{author}{\bibfnamefont{D.}~\bibnamefont{Stroud}},
  \bibinfo{journal}{Solid State Physics} \textbf{\bibinfo{volume}{46}},
  \bibinfo{pages}{147} (\bibinfo{year}{1992});
%
\bibinfo{author}{\bibfnamefont{A.}~\bibnamefont{Sihvola}},
  \emph{\bibinfo{title}{Electromagnetic mixing formulas and applications}},
  vol.~\bibinfo{volume}{47} of \emph{\bibinfo{series}{IEE Electromagnetic Waves
  Series}} (\bibinfo{publisher}{The Institute of Electrical Engineers},
  \bibinfo{address}{London}, \bibinfo{year}{1999}).

\bibitem[{\citenamefont{Bergman}(1982)}]{bergman1982}
\bibinfo{author}{\bibfnamefont{D.~J.} \bibnamefont{Bergman}},
  \bibinfo{journal}{Phys. Rep.}
  \textbf{\bibinfo{volume}{43}}(\bibinfo{number}{9}), \bibinfo{pages}{377}
  (\bibinfo{year}{1978});
\bibinfo{author}{\bibfnamefont{D.~J.} \bibnamefont{Bergman}},
  \bibinfo{journal}{Phys. Rev. Let.}
  \textbf{\bibinfo{volume}{44}}(\bibinfo{number}{19}), \bibinfo{pages}{1285}
  (\bibinfo{year}{1980});
\bibinfo{author}{\bibfnamefont{D.~J.} \bibnamefont{Bergman}},
  \bibinfo{journal}{Ann Phys} \textbf{\bibinfo{volume}{138}},
  \bibinfo{pages}{78} (\bibinfo{year}{1982}).
%
%

\bibitem[{\citenamefont{Milton}(1981{\natexlab{a}})}]{Milton1981}
\bibinfo{author}{\bibfnamefont{G.~W.} \bibnamefont{Milton}},
  \bibinfo{journal}{J Appl Phys} \textbf{\bibinfo{volume}{52}},
  \bibinfo{pages}{5286} (\bibinfo{year}{1981}{\natexlab{a}});
%
\bibinfo{author}{\bibfnamefont{G.~W.} \bibnamefont{Milton}},
  \bibinfo{journal}{J Appl Phys}
  \textbf{\bibinfo{volume}{52}}(\bibinfo{number}{8}), \bibinfo{pages}{5294}
  (\bibinfo{year}{1981}{\natexlab{b}}).

\bibitem[{\citenamefont{Golden and Papanicolaou}(1985)}]{Golden1}
\bibinfo{author}{\bibfnamefont{K.}~\bibnamefont{Golden}} \bibnamefont{and}
  \bibinfo{author}{\bibfnamefont{G.}~\bibnamefont{Papanicolaou}},
  \bibinfo{journal}{J Stat Phys}
  \textbf{\bibinfo{volume}{40}}(\bibinfo{number}{4/5}), \bibinfo{pages}{655}
  (\bibinfo{year}{1985}).

\bibitem[{\citenamefont{Ghosh and Fuchs}(1988)}]{GhoshFuchs}
\bibinfo{author}{\bibfnamefont{K.}~\bibnamefont{Ghosh}} \bibnamefont{and}
  \bibinfo{author}{\bibfnamefont{R.}~\bibnamefont{Fuchs}},
  \bibinfo{journal}{Phys Rev B}
  \textbf{\bibinfo{volume}{38}}(\bibinfo{number}{8}), \bibinfo{pages}{5222}
  (\bibinfo{year}{1988});
%
\bibinfo{author}{\bibfnamefont{A.~V.} \bibnamefont{Goncharenko}},
  \bibinfo{author}{\bibfnamefont{V.~Z.} \bibnamefont{Lozovski}},
  \bibnamefont{and} \bibinfo{author}{\bibfnamefont{E.~F.}
  \bibnamefont{Venger}}, \bibinfo{journal}{Opt Comm}
  \textbf{\bibinfo{volume}{174}}, \bibinfo{pages}{19} (\bibinfo{year}{2000});
%
\bibinfo{author}{\bibfnamefont{A.~V.} \bibnamefont{Goncharenko}},
  \bibinfo{journal}{Phys Rev E}
  \textbf{\bibinfo{volume}{68}}, \bibinfo{pages}{041108}
  (\bibinfo{year}{2003}).

\bibitem[{\citenamefont{Debye}(1945)}]{Debye1945}
\bibinfo{author}{\bibfnamefont{P.}~\bibnamefont{Debye}},
  \emph{\bibinfo{title}{Polar Molecules}} (\bibinfo{publisher}{Dover
  Publications}, \bibinfo{address}{New York}, \bibinfo{year}{1945}).

\bibitem[{\citenamefont{Tuncer and Guba{\'n}ski}(2001)}]{Tuncer2000b}
\bibinfo{author}{\bibfnamefont{E.}~\bibnamefont{Tuncer}} \bibnamefont{and}
  \bibinfo{author}{\bibfnamefont{S.~M.} \bibnamefont{Guba{\'n}ski}},
  \bibinfo{journal}{IEEE Trans Diel El Insul}
  \textbf{\bibinfo{volume}{8}}, \bibinfo{pages}{310} (\bibinfo{year}{2001}).


\bibitem[{\citenamefont{Onsager}(1936)}]{Onsager}
\bibinfo{author}{\bibfnamefont{L.}~\bibnamefont{Onsager}},
  \bibinfo{journal}{J Am Chem Soc}
  \textbf{\bibinfo{volume}{58}}, \bibinfo{pages}{1486} (\bibinfo{year}{1936});
\bibinfo{author}{\bibfnamefont{H.}~\bibnamefont{Fr{\"o}hlich}},
  \emph{\bibinfo{title}{Theory of Dielectrics; Dielectric constant and
  dielectric loss}} (\bibinfo{publisher}{Oxford Science Publications},
  \bibinfo{address}{Oxford}, \bibinfo{year}{1958}), \bibinfo{edition}{2nd} ed;
\bibinfo{author}{\bibfnamefont{U.}~\bibnamefont{Fano}},
  \bibinfo{journal}{Phys Rev} \textbf{\bibinfo{volume}{118}},
  \bibinfo{pages}{451} (\bibinfo{year}{1960});
\bibinfo{author}{\bibfnamefont{S.~E.} \bibnamefont{Schnatterly}}
  \bibnamefont{and} \bibinfo{author}{\bibfnamefont{C.}~\bibnamefont{Tarrio}},
  \bibinfo{journal}{Revi Mod Phys} \textbf{\bibinfo{volume}{64}},
  \bibinfo{pages}{619} (\bibinfo{year}{1992});
\bibinfo{author}{\bibfnamefont{B.~K.~P.} \bibnamefont{Scaife}},
  \emph{\bibinfo{title}{Principles of Dielectrics}} (\bibinfo{publisher}{Oxford
  Science Publications}, \bibinfo{year}{1998}).

\bibitem[{\citenamefont{Havriliak and Negami}(1966)}]{HN}
\bibinfo{author}{\bibfnamefont{S.}~\bibnamefont{Havriliak}} \bibnamefont{and}
  \bibinfo{author}{\bibfnamefont{S.}~\bibnamefont{Negami}},
  \bibinfo{journal}{J Polym Sci: Part C}
  \textbf{\bibinfo{volume}{14}}, \bibinfo{pages}{99} (\bibinfo{year}{1966}).

\bibitem[{\citenamefont{Davidson and Cole}(1951)}]{CD}
\bibinfo{author}{\bibfnamefont{D.~W.} \bibnamefont{Davidson}} \bibnamefont{and}
  \bibinfo{author}{\bibfnamefont{R.~H.} \bibnamefont{Cole}},
  \bibinfo{journal}{J Chem Phys} \textbf{\bibinfo{volume}{19}},
  \bibinfo{pages}{1484} (\bibinfo{year}{1951}).

\bibitem[{\citenamefont{Cole and Cole}(1941)}]{CC}
\bibinfo{author}{\bibfnamefont{K.~S.} \bibnamefont{Cole}} \bibnamefont{and}
  \bibinfo{author}{\bibfnamefont{R.~H.} \bibnamefont{Cole}},
  \bibinfo{journal}{J Chem Phys} \textbf{\bibinfo{volume}{9}},
  \bibinfo{pages}{341} (\bibinfo{year}{1941}).

\bibitem[{\citenamefont{Jonscher}(1983)}]{Jonscher1983}
\bibinfo{author}{\bibfnamefont{A.~K.} \bibnamefont{Jonscher}},
  \emph{\bibinfo{title}{Dielectric Relaxation in Solids}}
  (\bibinfo{publisher}{London: Chelsea Dielectric}, \bibinfo{address}{London},
  \bibinfo{year}{1983});
\bibinfo{editor}{\bibfnamefont{J.~R.} \bibnamefont{Macdonald}}, ed.,
  \emph{\bibinfo{title}{Impedance Spectroscopy}} (\bibinfo{publisher}{John
  Wiley \& Sons}, \bibinfo{address}{New York}, \bibinfo{year}{1987});
%
\bibinfo{author}{\bibfnamefont{J.~R.} \bibnamefont{Macdonald}},
  \bibinfo{journal}{J Comp Phys}
  \textbf{\bibinfo{volume}{157}}, \bibinfo{pages}{280} (\bibinfo{year}{2000});
%
\bibinfo{author}{\bibfnamefont{J.~R.} \bibnamefont{Macdonald}},
  \bibinfo{journal}{J Non-Cryst Solids}
  \textbf{\bibinfo{volume}{212}}, \bibinfo{pages}{95} (\bibinfo{year}{1997});
%
\bibinfo{author}{\bibfnamefont{J.~R.} \bibnamefont{Macdonald}}
  \bibnamefont{and} \bibinfo{author}{\bibfnamefont{L.~D.}
  \bibnamefont{Potter Jr.}}, \bibinfo{journal}{Solid State Ionics}
  \textbf{\bibinfo{volume}{24}}(\bibinfo{number}{1}), \bibinfo{pages}{61}
  (\bibinfo{year}{1987}). 
%


\bibitem[{\citenamefont{Garnett}(1904)}]{Maxwell_Garnett}
\bibinfo{author}{\bibfnamefont{J.~C.~M.} \bibnamefont{Garnett}},
  \bibinfo{journal}{Phil Trans Royal Soc London A}
  \textbf{\bibinfo{volume}{203}}, \bibinfo{pages}{385} (\bibinfo{year}{1904});
\bibinfo{author}{\bibfnamefont{O.}~\bibnamefont{Levy}} \bibnamefont{and}
  \bibinfo{author}{\bibfnamefont{D.}~\bibnamefont{Stroud}},
  \bibinfo{journal}{Phys Rev B}
  \textbf{\bibinfo{volume}{56}}(\bibinfo{number}{13}), \bibinfo{pages}{8035}
  (\bibinfo{year}{1997}).

\bibitem[{\citenamefont{Barrera and Fuchs}(1995)}]{BarreraPRB1995}
\bibinfo{author}{\bibfnamefont{R.~G.} \bibnamefont{Barrera}} \bibnamefont{and}
  \bibinfo{author}{\bibfnamefont{R.}~\bibnamefont{Fuchs}},
  \bibinfo{journal}{Phys Rev B} \textbf{\bibinfo{volume}{52}},
  \bibinfo{pages}{3256} (\bibinfo{year}{1995});
\bibinfo{author}{\bibfnamefont{J.}~\bibnamefont{Monecke}},
  \bibinfo{journal}{Phys Rev B} \textbf{\bibinfo{volume}{55}},
  \bibinfo{pages}{7515} (\bibinfo{year}{1997});
\bibinfo{author}{\bibfnamefont{R.~O.} \bibnamefont{Kuzian}},
  \bibinfo{author}{\bibfnamefont{R.}~\bibnamefont{Hayn}}, \bibnamefont{and}
  \bibinfo{author}{\bibfnamefont{A.~F.} \bibnamefont{Barabanov}},
  \bibinfo{journal}{Physical Review B} \textbf{\bibinfo{volume}{68}},
  \bibinfo{pages}{195106} (\bibinfo{year}{2003});
\bibinfo{author}{\bibfnamefont{M.~I.} \bibnamefont{Stockman}},
  \bibinfo{author}{\bibfnamefont{D.~J.} \bibnamefont{Bergman}},
  \bibnamefont{and}
  \bibinfo{author}{\bibfnamefont{T.}~\bibnamefont{Kobayashi}},
  \bibinfo{journal}{Physical Review B} \textbf{\bibinfo{volume}{69}},
  \bibinfo{pages}{054202} (\bibinfo{year}{2004}).


\end{thebibliography}

\end{document}